\documentclass[twocolumn]{aa} 
\usepackage{aalongtable}        

\usepackage{graphicx}
\usepackage{aalongtable}        
\usepackage{natbib}             
\bibpunct{(}{)}{;}{a}{}{,}      
\usepackage{txfonts}

\begin{document}
\title{FUNDPAR: A program for Deriving Fundamental Parameters from Equivalent Widths}

\author{C. Saffe \inst{1}}

\institute{Instituto de Ciencias Astron\'omicas, de la Tierra y del Espacio (ICATE), C.C 467,
5400, San Juan, Argentina. Members of the Carrera del Investigador Cient\'{\i}fico, CONICET,
Consejo Nacional de Investigaciones Cient\'{i}ficas y T\'{e}cnicas de la Rep\'{u}blica Argentina}
\date{Received xx/xx/xx; accepted xx/xx/xx}


\abstract
{}
{We implemented a fortran code that determine fundamental parameters of solar
type stars from a list of Fe line equivalent widths.
The solution should verify 3 conditions in the standard method:
ionization equilibrium, excitation equilibrium and independence between metallicity
and equivalent widths. We added the condition that the input metallicity
of the model atmosphere should be similar to the output metallicity derived with
equivalent widths.}
{Solar-scaled Kurucz model atmospheres with NEWODF opacities are
calculated with an independent program.
Parameter files control different details, such as the
mixing-length parameter, the overshooting, the damping of the lines and the weight
factors in the definition of the $\chi^2$ function.}
{FUNDPAR derive the uncertainties following 2 methods: the criteria of Gonzalez \&
Vanture (1998) and the dispersion using the $\chi^2$ function.
The code use the 2009 version of the MOOG program.
The results derived with FUNDPAR are in agreement with previous determinations in
the literature.
In particular we obtained the fundamental parameters of 58 exoplanet host stars.
The program is freely available from the web\footnote{http://icate-conicet.gob.ar/saffe/fundpar/}.}
{}

\maketitle

\section{Introduction}

Different methods have been used in the literature
to derive fundamental parameters and metallicities of solar type stars.
For instance, some studies begin with a photometric estimation of temperature 
and gravity and then derive the metallicity using equivalent widths.
Useful codes such as WIDTH9 \citep{kurucz93,kurucz95} or
BLACKWEL\footnote{http://www1.appstate.edu/dept/physics/spectrum/spectrum.html}
(an implementation of the Blackwell method) are widely used in the
literature \citep[e.g. ][]{saffe-blackwell}. However the temperature and
gravity are ussually considered as fixed parameters.
The observed stellar spectra could be compared with a grid of previously
calculated synthetic spectra \citep[e.g. ][]{fischer-valenti05,saffe-vega} to
determine fundamental parameters.
However in this case the instrumental broadening should be taken into account
and the rotational velocity could be considered as another independent parameter.

Recently, \citet{sousa10} derived an effective temperature calibration based
on line equivalent width ratios of different absorption lines.
Also the equivalent widths of Fe lines could be used to determine the parameters
of solar type stars.
The solution should verify three conditions in the standard method:
{[FeI/H]$=$[FeII/H]} (i.e. ionization equilibrium), independence of the metallicity
with the excitation potential (i.e. excitation equilibrium) and with respect to
the equivalent widths.  This method have been applied to solar type stars, for instance,
in the determination of metallicity of stars with and without low mass companions or
exoplanets \citep[e.g. ][]{gonzalez97,gonzalez98,gonzalez99,santos00,gonzalez-laws00,
gonzalez01,laws-gonzalez01,laws03}.
The process estimate initially the fundamental parameters
(T$_{eff}$, $log g$, [Fe/H] and $\xi$, the microturbulence velocity).
This information is used in the ATLAS \citep{kurucz93,kurucz95} program to derive
a model atmosphere in LTE (local thermodynamic equilibrium).
The model atmosphere with the measured equivalent widths in the spectra, are
introduced in the code MOOG\footnote{http://verdi.as.utexas.edu/moog.html}
to derive a new metallicity.
If the mentioned conditions are not satisfied, the process is restarted
using new fundamental parameters calculated with the downhill method.

In this contribution, we present the fortran code FUNDPAR (and their
complement atlas.launcher) that reproduce the method explained. The programs
are available from the web\footnote{http://icate-conicet.gob.ar/saffe/fundpar/},
including detailed installation instructions and some technical details such
as the format of the input/output files (install.txt).
As an example of practical use, we derived the fundamental parameters of 58
main sequence exoplanet host stars and verified the metal-rich nature of the
group. The values derived are in agreement with previous determinations from
literature.

In the section 2 we show the general idea and the logic of the program.
The procedure of minimization of the $\chi^2$ function is detailed in the section 3.
The estimation of the uncertainties in the parameters and the comparison
with literature, are showed in the sections 4 and 5, respectively.
Finally we present some concluding remarks in the section 6.

\section{The logic of the program}

The algorithm is organized in one main procedure and a number of sub-programs
with specific functions. The iterative process begins in one starting point
in a 4D-space, where the variables are (T$_{eff}$, $log g$, [Fe/H], $\xi$).
This starting point could be optionally given in the first line of
the input file (see the file install.txt for details of input/output files).
If this point is unknown, the program adopt (5000 K, 4.00 dex,
-0.10 dex, 1.00 km/s) to begin the iteration. For example, \citet{santos00}
estimate T$_{eff}$ and $log g$ from the uvby photometry and the calibration
of \citet{olsen84}, while [Fe/H] and $\xi$ are estimated following
\citet{schuster-nissen89} and \citet{edvar93}, respectively.

Then for each iteration step, i.e. for each group of 4 parameters
(T$_{eff}$, $log g$, [Fe/H], $\xi$), the program should:
\\ \\a) Generate an appropriate ATLAS model atmosphere (through the atlas.launcher
program). First, the code should select from the Kurucz grid the closer model atmosphere
in the parameter space to those requested and take this as initial input model.
Then the program execute ATLAS9 to derive the final model.
\\ \\b) Transform the model atmosphere in a format readable by MOOG
(subroutine kurucz2moog). The format of the model atmosphere used
by MOOG is not exactly the Kurucz model and should be rewritten
accordingly. 
\\ \\c) Call the MOOG program. In this point, the program takes the file
containing the equivalent widths of FeI and FeII lines of the star
and the model atmosphere as input for the MOOG program.
The MOOG program is executed using a driver called ''abfind''
which is selected for the abundance determination.
\\ \\d) Read the new metallicity values calculated by MOOG (subroutine
rmr, read-moog-results) and finally,
\\ \\e) Determine the value of the $\chi^2$ function.
This step will be explained below. \\

We take into account the conditions mentioned in the introduction in a
variable called $\chi^2$. We adopt for $\chi^2$ the expression
{$\chi^2 = w_{1} c_{1}^2 + w_{2} c_{2}^2 + w_{3} c_{3}^2 + w_{4} c_{4}^2$,}
where {w$_{1}$,...,w$_{4}$} are considered weight factors (w$_{i}>=$0),
c$_{1}$ and c$_{2}$ are the slopes in the plots of [Fe/H] vs. {$log_{10}(W/\lambda)$}
(logarithm of the reduced equivalent width)
and [Fe/H] vs. excitation potential, {c$_{3}=$[FeI/H]-[FeII/H]}, and c$_{4}$ is the difference
between the input ATLAS metallicity (step a) and the resulting metallicity using
equivalent widths (step d). 
We added explicity the fourth condition: the input metallicity of the model
atmosphere should be similar to the output metallicity derived with equivalent widths
i.e. the term with c$_{4}$.
Then, the 4 conditions are quantified in the $\chi^2$
function: the solution correspond to the minimum value of $\chi^2$.

The user is free to modify the values of the weights {w$_{1}$,...,w$_{4}$} under their
own criteria. However we show a brief example estimating aproximately the values of the weights.
Adopting $\chi^2=$1 as the limit case of a solution, each condition
contribute, for example, with 0.25 to the sum {$\chi^2=$ 0.25+...+0.25}. 
In this case the 4 conditions are taken equally important within $\chi^2$, which is not
always true.
In the plot of abundance vs. excitation potential, we accept a maximum slope,
for instance, of {$c_{1}\sim$ 0.015/4 dex/eV}, taking a difference of {$\sim$0.015 dex}
in abundance for a total range of $\sim$4 eV in the excitation potential of Fe lines.
Then in the limit case, {$w_{1} c_{1}^2 \sim$ 0.25} and thus {$w_{1}\sim$ 18000 eV$^2$/dex$^2$}.
The units of {$w_{1}$} are forced to obtain the product {$w_{1} c_{1}^2$} without units.
In the plot of abundance vs. {log${_{10}}$(W/$\lambda$)} (where W and $\lambda$ are the
equivalent width and wavelength in \AA, respectively), we accept a maximum slope (for example)
of {$c_{2}\sim$ 0.015/1.5}, taking a difference of 0.015 dex in abundance for a
range of $\sim$1.5 in the log${_{10}}$(W/$\lambda$) of Fe lines. Then,
{$w_{2} c_{2}^2 \sim$ 0.25} and thus {$w_{2}\sim$ 2500}.
For the third condition, {$c_{1}=$[FeI/H]-[FeII/H]} and we adopt a maximum difference
of 0.015 dex. Then, {$w_{3} c_{3}^2 \sim$ 0.25} and thus {$w_{3}\sim$ 1100 dex$^{-2}$}.
Similarly, for $w_{4}$ result {$w_{4}\sim$ 1100 dex$^{-2}$}. 
In this estimation the weights {w$_{1}$,...,w$_{4}$} resulted 18000 eV$^2$/dex$^2$, 2500,
1100 dex$^{-2}$ and 1100 dex$^{-2}$, respectively, for a solution in which
the 4 conditions contribute equally with 0.25 to the $\chi^2$ function in the
limit case of $\chi^2=$1. In this example those solutions with $\chi^2>$1 do not
verify the four conditions.

It is probably that the user have their own criteria adopting the values of {w$_{1}$,...,w$_{4}$},
instead of the example explained the previous paragraph.
The user is free to modify the values of {w$_{1}$,...,w$_{4}$} (file fundpar.par) and this could
result in more (or less) restrictive conditions.
The code use this values to define $\chi^2$ and then search the minimum of the function.
Then, the user should read the values of the slopes and metallicities in the output files
to verify if the 4 conditions are satisfied.
The values of {w$_{1}$,...,w$_{4}$} previously showed seems to verify in practice the
requeriments of minimization and verification of the 4 conditions.
In the Table \ref{fundpar.par} we show a sample of the file fundpar.par where the 
weights could be modified. Other parameters will be explained in the next sections.

\begin{table*}
\center
\caption{Sample of the fundpar.par file which determine the weights of the $\chi^2$function.}
\label{fundpar.par}
\begin{tabular}{llrl}
\hline
t1&=&   200.00 & \# Characteristic length scale of Teff (K) \\
t2&=&     0.10 & \# Characteristic length scale of Logg (dex) \\
t3&=&     0.10 & \# Characteristic length scale of [Fe/H] (dex)\\
t4&=&     0.10 & \# Characteristic length scale of Xita (km/s)\\
w1&=& 25000.00 & \# Chi2 Weight factor \\
w2&=&  2500.00 & \# Chi2 Weight factor \\
w3&=&  1111.00 & \# Chi2 Weight factor \\
w4&=&  1111.00 & \# Chi2 Weight factor \\
\hline
\end{tabular}
\end{table*}

Following the definition, $\chi^2$ could be considered as a function that depends of the
fundamental parameters {$\chi^2$=$\chi^2$(T$_{eff}$,$log g$,[Fe/H],$\xi$)}. If $\chi^2$ is not
minimum, the algorithm should determine the next
set of 4 possible values. These new variables are used in another
iteration step (following the steps a to e) to derive a new model atmosphere,
metallicity and finally a new value of $\chi^2$.
The algorithm that determine the next group of 4 parameters is the
downhill method, explained in the next section.

In the Table \ref{input.output} we show a list of the input and output files used
by FUNDPAR. The format of the input/output files is detailed in the file install.txt.
There are two main directories (datain and dataout) containing the input and output
files of the stars. 
The equivalent widths should be stored in separate files (one file by star), and the 
names of these files should be listed within another file called filenames.txt.
The files atlas.par, batch.par and fundpar.par determine
the value of some parameters used in the model calculation and abundance determination
and will be explained in the next sections. After the execution of FUNDPAR,
there are three output files by star: the ATLAS model atmosphere of the solution
and two output files directly from the MOOG abundance determination.
The file output1.screen contain information similar to the screen and
output2.results list the final parameters and their uncertainties.

\begin{table*}
\center
\caption{List of the input and output files used by FUNDPAR.}
\label{input.output}
\begin{tabular}{lll}
Filename & Directory & Comment \\
\hline \hline
Input Files:\\
\hline
filenames.txt & datain &  List of files containing the equivalent widths. \\
hd001.txt     & datain &  Equivalent widths of the star HD 001 \\
...           &  ...   &  ... \\
hd999.txt     & datain &  Equivalent widths of the star HD 999 \\
\hline
Parameter Files:\\
\hline
atlas.par     & atlas.launcher &  Parameters used by ATLAS \\
batch.par     & FUNDPAR & Parameters used by MOOG \\
fundpar.par   & FUNDPAR & Parameters used by FUNDPAR \\
\hline
Output Files:\\
\hline
hd001.mod      & dataout & Model atmosphere of HD 001 from ATLAS9\\
hd001.out1.txt & dataout & Output file 1 from MOOG \\
hd001.out2.txt & dataout & Output file 2 from MOOG \\
...            & ...     & ... \\
output1.screen & dataout & Similar data to the screen\\
output2.results& dataout & Table of results for all the stars \\
\hline
\end{tabular}
\end{table*}

\section{The downhill simplex method: minimization of $\chi^2$}

In this section we briefly review the minimization procedure of $\chi^2$ as a
function of 4 independent variables, using a Numerical Recipe's routine called 
{\it{amoeba}} \citep{press92}. The downhill simplex method is due to \citet{nelder-mead65}
and requires only function evaluations, not derivatives.
A simplex could be considered as a geometrical figure of N+1 vertices
in a N-dimensional space (in our case, N=4).
Taking any vertice as the origin, then the 4 other points define possible
vector directions in the 4-dimensional volume.

The downhill simplex method start with a group of N+1 i.e. 5 vertices rather
than a single point or vertice. These vertices are desplaced in a
characteristic length scale of the problem. In our case, $\chi^2$ is initially calculated
adopting displacements of 200 K, 0.1 dex, 0.1 dex and 0.1 km/s
for the variables T$_{eff}$, $log g$, [Fe/H] and $\xi$, respectively.
The characteristic lengths could be modified in the file fundpar.par (see Table \ref{fundpar.par}).
FUNDPAR execute an initial calculation of $\chi^2$ at the vertices of the simplex,
previous to the iteration process.
Then, the downhill method takes a series of steps, ussually moving the
highest point of the simplex i.e. where $\chi^2$ is maximum. Succesive steps could be
visualized as reflections, expansions and contractions of the 4-dimensional
object. When the simplex found a valley, it contracts itself down the valley.
An appropriate sequence of these steps will converge to the minimum of $\chi^2$.

As explained by \citet{press92}, the termination criteria is ussually delicate
in the minimization process.
The program require that the decrease in the function value in the terminating
step be fractionally smaller that some tolerance (variable ftol within the
amoeba subroutine). The method explained exactly follows \citet{nelder-mead65}.
However, in practice we found that for some stars the routine converge in a few iteration steps
to a solution with $\chi^2>>$1. In such cases the code decrease the 
tolerance and continue the iteration process from the last point.
We verified that it is not necessary in this case to restart again all the process, because
the effect of modify the tolerance determine only in which iteration the program stops.
We also note that decrease the tolerance do not guarantee the convergence of the solution:
we are only modifying the termination step or the termination criteria.

Using the adopted values of {w$_{1}$,...,w$_{4}$}, the program ussually takes less
than $\sim$200 iterations to reach a solution. 
The number of iterations is a known problem of the downhill method.
If the code reach a solution with $\chi^2>$1, then the program restart the iteration process
using the last solution as a new initial condition.
The restart is recomended by \citet{press92} in the downhill method to eliminate a probable
local minimum of the function.
The code restart the iteration process only once. Also the user could manually restart
again all the process using the last solution, for instance, as new initial condition
and FUNDPAR will try to search a solution with a smaller $\chi^2$.

\section{Derivation of the model atmospheres} 

Together with the {FUNDPAR} code, we provide another independent program
called atlas.launcher, which prepare and execute the Kurucz's ATLAS9 in LTE.
The input of atlas.launcher are the fundamental parameters T$_{eff}$, $log g$,
[Fe/H] and $\xi$. The output is an ATLAS solar-scaled model atmosphere corresponding
to these parameters. The program use a parameter file called atlas.par that
determine different parameters used by ATLAS, such as the mixing-length
parameter (ussually taken as 1.25), and the overshooting weight parameter W,
defined in \citet{castelli97}. A list of files used by ATLAS (grid of precalculated
models, Rosseland and ODFs, and a sample of atlas.par) is presented in the file
install.txt.

Then, the model atmosphere and the equivalent widths are introduced in the MOOG program.
We instructed this code with the solar abundances of \citet{grevesse-sauval98} instead of
the original from \citet{anders-grevesse89}, except for Fe for which we adopted
{[Fe/H]=7.47 dex}. The NEWODF opacities use these abundance values \citep{castelli-kurucz03}.
MOOG use a parameter file (batch.par) where some options\footnote{For a complete
list of options, see the manual of the MOOG program, http://verdi.as.utexas.edu/moog.html}
could be modified, such as the molecular equilibrium and the van der Waals line
damping options.

Literature authors have different preferences in the choice of ATLAS and MOOG
parameters. The user is free to modify the parameters under their own criteria, and then
FUNDPAR will found the corresponding solution.
To give an idea of the sensivity of the results, we rederived the fundamental parameters
of 5 sample stars (HD 106252, HD 177830, HD 190228, HD 195019 and HD 202206) using different
combination of the parameters. In the Table \ref{combin.pars} we show the values of
the parameters adopted in different calculations A,...,E. The parameters showed are
the mixing-length parameter (ML), the overshooting parameter (W) and the line damping
option (D). We start the execution A adopting ML$=$1.25, W$=$1 and D$=$2.
In the execution B we switched off the overshooting, while in the execution C we
adopted a slightly higher ML. In the executions D and E we used different options
for the line damping.

\begin{table}
\center
\caption{Parameters adopted in different executions A,...,E for 5 sample stars.
(See text for details)}
\label{combin.pars}
\begin{tabular}{llll}
\hline
Run & ML & W & D \\
\hline
A & 1.25 & 1 & 2 \\
B & 1.25 & 0 & 2 \\
C & 1.32 & 1 & 2 \\
D & 1.25 & 1 & 1 \\
E & 1.25 & 1 & 0 \\
\hline
\end{tabular}
\end{table}

\begin{figure*}
\center
\includegraphics[width=60mm]{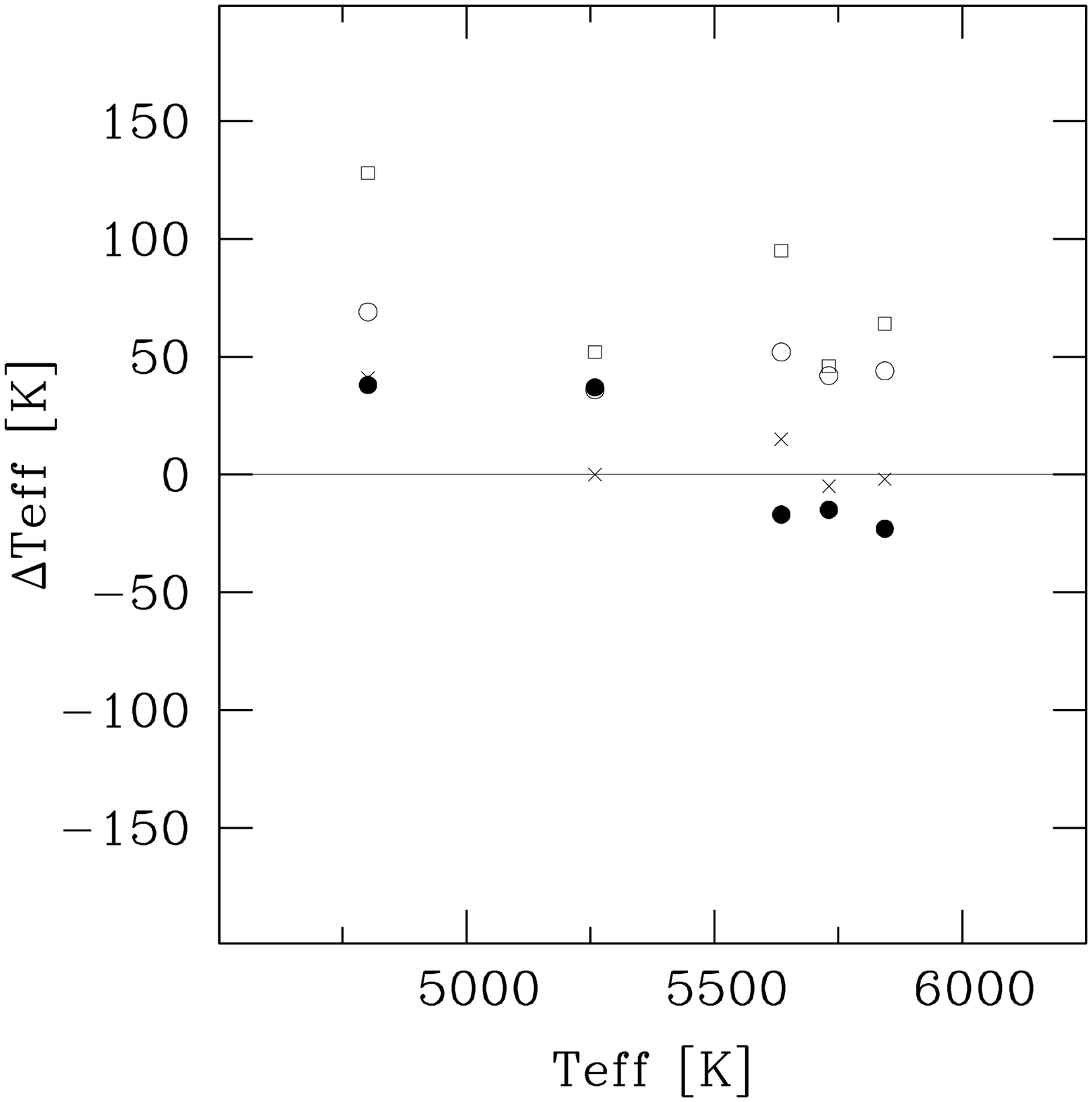}
\includegraphics[width=60mm]{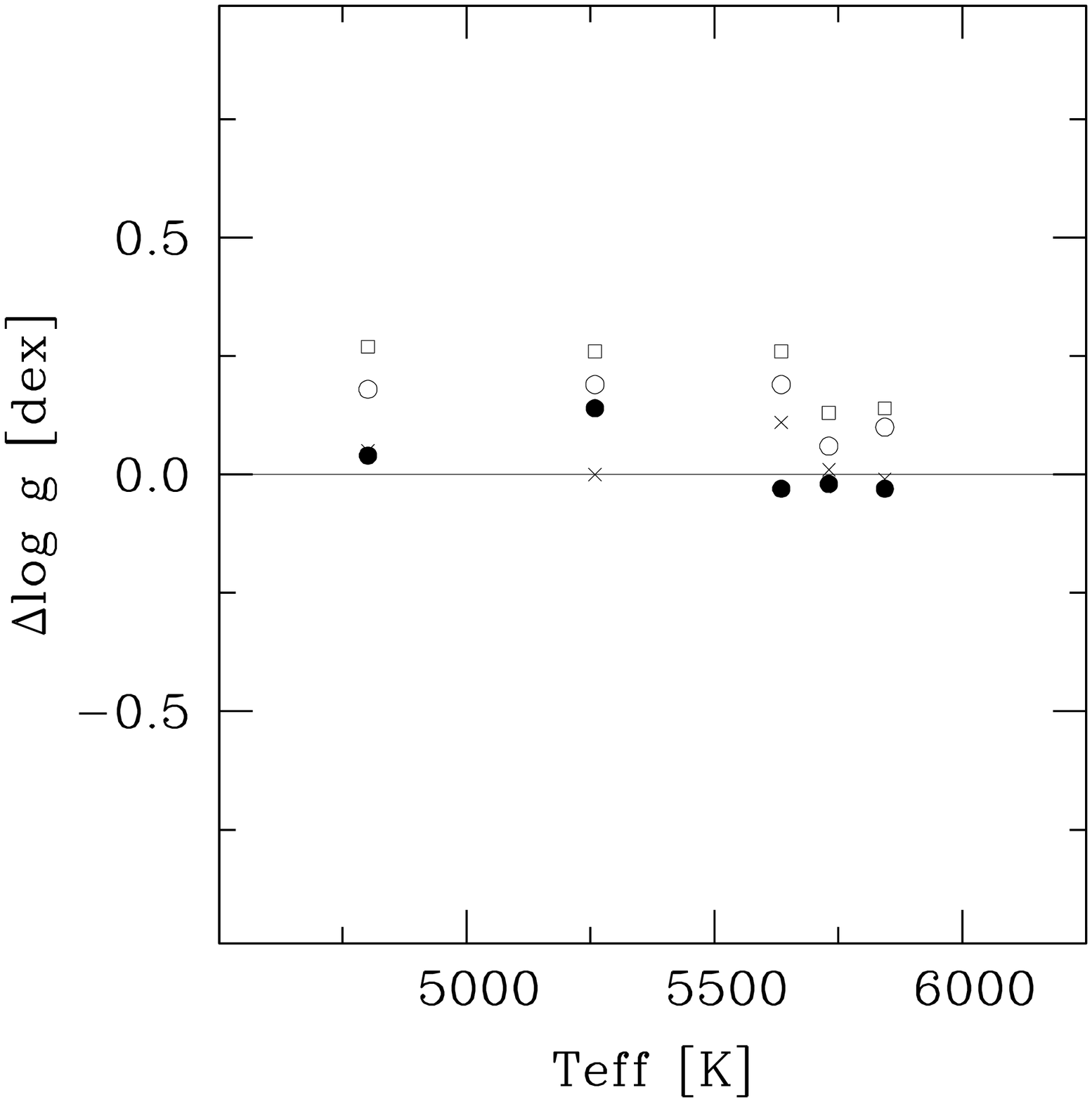}
\includegraphics[width=60mm]{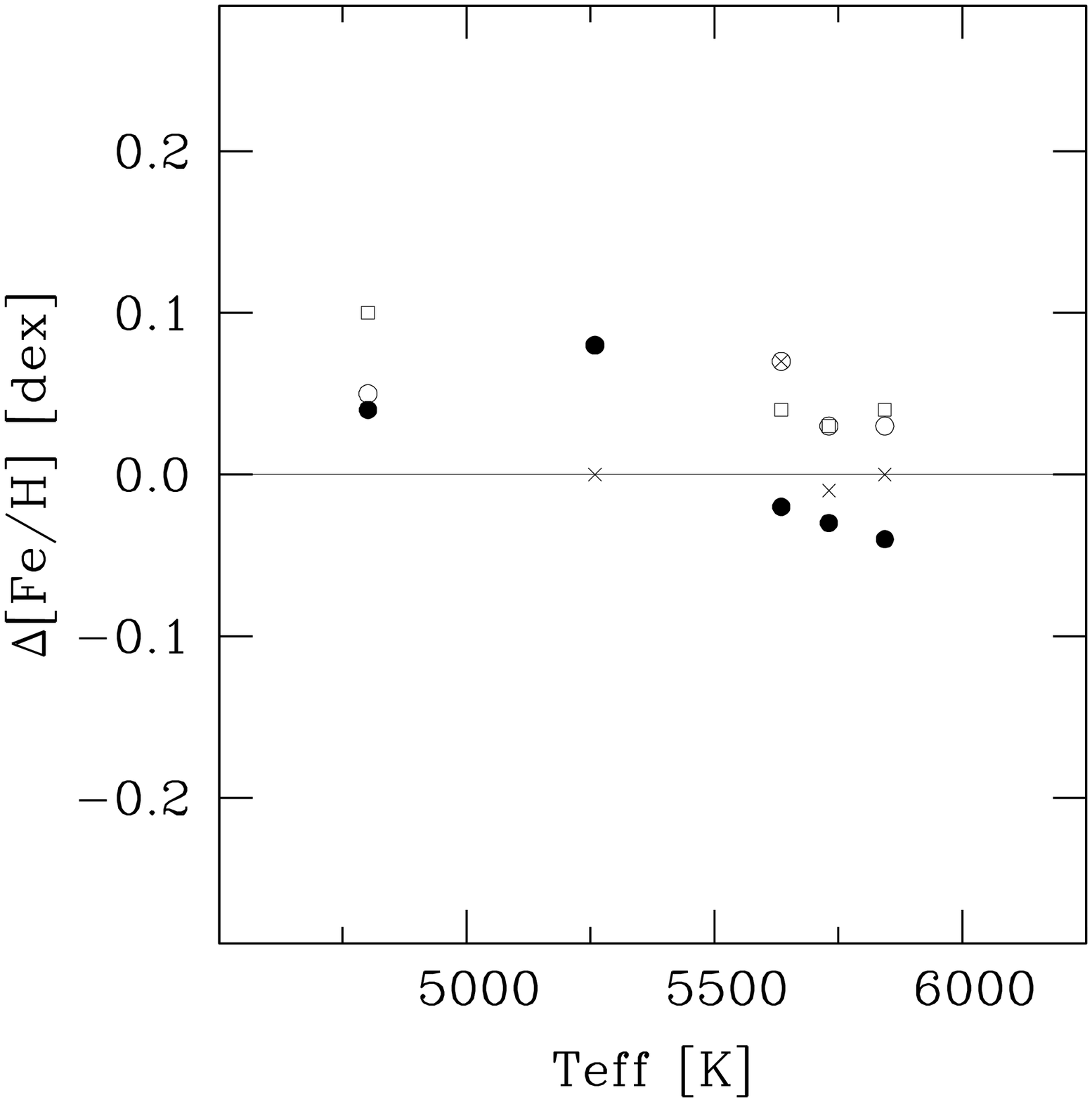}
\includegraphics[width=60mm]{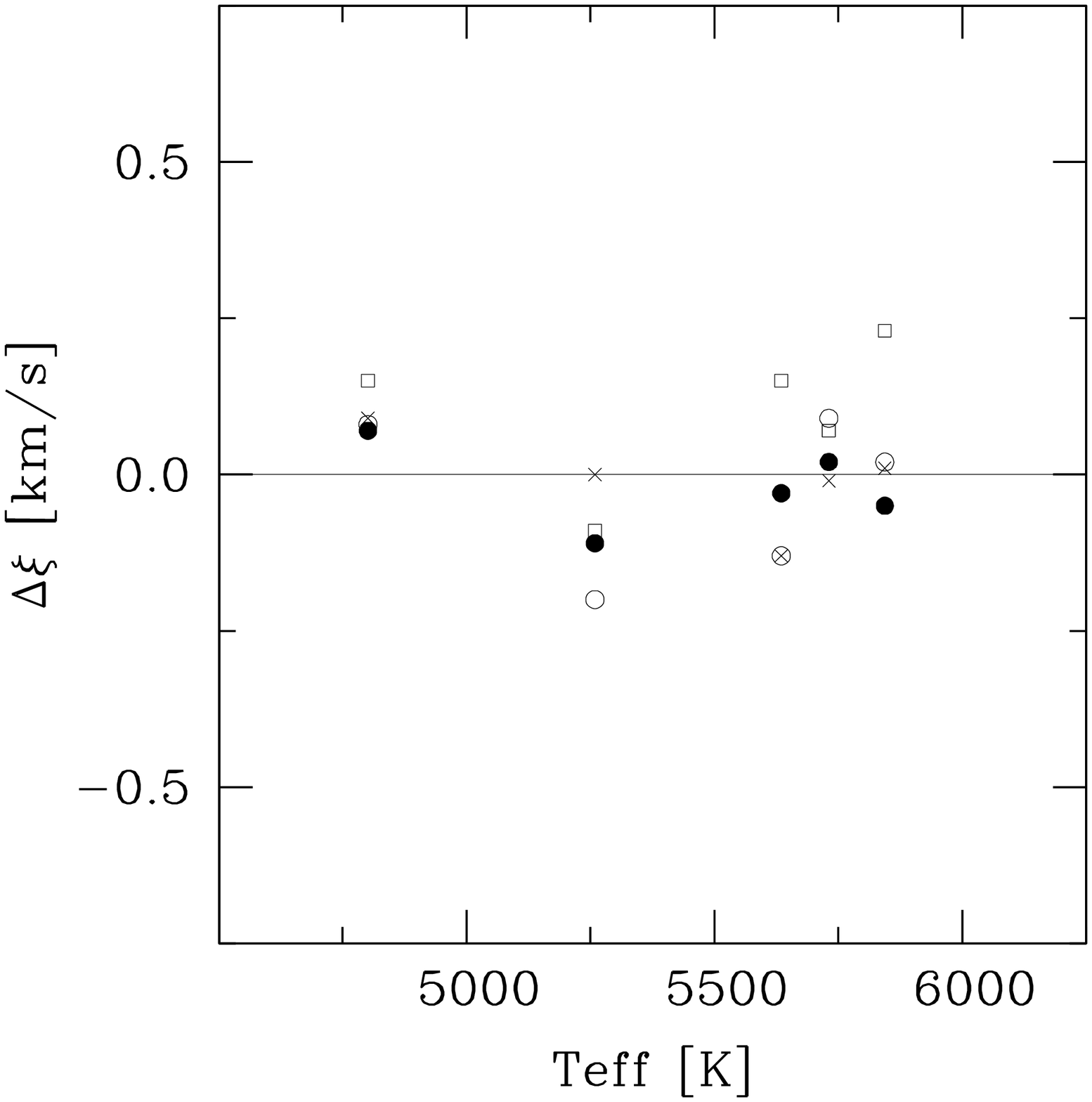}
\caption{Difference between the parameters derived in the A execution and the parameters
derived in the executions B, C, D and E. The differences are plotted vs. T$_{eff}$ only
for 5 stars. For instance, in the T$_{eff}$ panel we ploted $\Delta$T$_{eff}=$
T$_{eff}$(B)-T$_{eff}$(A) with filled circles. The symbols used in the panels are
filled circles (differences B-A), diagonal crosses (differences C-A), empty circles
(differences D-A) and squares (differences E-A).}
\label{fig.pars}
\end{figure*}

The variation in T$_{eff}$, log g, [Fe/H] and $\xi$ (adopting different
ML, W and D) is different from star to star. For instance, varying W 
some stars increase the T$_{eff}$ while others decrease the T$_{eff}$. Probably the
variation depends of the fundamental parameters of the stars. Then,
in the Figure \ref{fig.pars} we show the difference (called $\Delta$) between the
parameters derived in the A execution and the parameters derived in the executions
B, C, D and E. The differences are plotted vs. T$_{eff}$ only for the 5 stars
previously mentioned. All differences are derived with respect to the A execution.
The symbols used in the panels are filled circles (results B-A),
diagonal crosses (results C-A), empty circles (results D-A) and squares (results E-A).
The panels seems to show a slight tendence with T$_{eff}$.
Switching on/off the overshooting (B execution), $\Delta$T$_{eff}$, $\Delta$log g and
$\Delta$[Fe/H] seems to show a tendence with T$_{eff}$. Increasing ML from 1.25 to 1.32
(C execution), we find no clear tendences and the differences are small.
Using the damping option 1 instead of 2 (D execution), T$_{eff}$, logg and [Fe/H]
increased an average of $\sim$48 K, $\sim$0.14 dex and $\sim$0.05 dex,
respectively. Using the damping option 0 (E execution), T$_{eff}$, logg, [Fe/H] and
$\xi$ increased an average of $\sim$70 K, $\sim$0.20 dex, $\sim$0.06 dex and
$\sim$0.1 km/s, respectively. We caution that these preliminar differences have been
derived using only 5 stars. A higher number of stars is needed to properly determine
how the parameters ML, W and D modify the derived fundamental parameters.

\section{Uncertainty estimation of the derived parameters}

The uncertainties in the atmospheric parameters were estimated following
\citet{gonzalez-vanture98}. The uncertainty in $\xi$ was determined from
the standard deviation in the slope of the least-squares fit of abundance vs.
reduced equivalent width.
Then, the dispersion in effective temperature was determined from the uncertainty
in the slope of the least-squares fit of abundance vs. excitation potential,
in addition to the uncertainty in the slope due to the uncertainty in $\xi$.
The uncertainty in the Fe abundance was derived combining the uncertainties in T$_{eff}$,
$\xi$ and the scatter of the individual FeI abundances (standard deviation of the
mean), all added in quadrature. In calculating the uncertainty in $log g$, we
include the contribution from the uncertainty in T$_{eff}$ in addition to the scatter
in the FeII line abundances.

We derive another estimation of the uncertainty of the solution using $\chi^2$.
As we explained previously, we selected the values of the weights such that solutions
with $\chi^2>1$ do not verify the 4 conditions.
Then, we adopt the size of the region $\chi^2=$1 as another estimation of the uncertainty
of the solution. FUNDPAR store a record of the points with $\chi^2<$1 and use them to
estimate the size of the region. The range
of the values of T$_{eff}$, $log g$, [Fe/H] and $\xi$ are showed in the results
with the number n of solutions with $\chi^2<$1. These dispersion values should be 
taken with caution if n is small. 
This kind of uncertainty is comparable to those derived by the criteria of
\citet{gonzalez-vanture98}. In the next section we present a histogram comparing
both uncertainties for a group of exoplanet host stars. 
The user is free to select between them, or the maximum value, for instance.

{FUNDPAR} include the uncertainty estimation of the parameters and they are showed in
the ouput files. The amount of the uncertainty depend on many factors, such as
the number of lines involved (which is ussually low for FeII), the measured equivalent
widths and the laboratory data of the spectral lines. The log gf and excitation
potential of the lines are important because many derived values (Fe abundances,
the slopes of abundance vs. equivalent widths and vs. excitation potential, etc.)
depends on these quantities.
We note that the inclusion of a line which (for one reason or another) result in
an abundance very different from the average, could modify the results and/or
increase the uncertainty of the parameters. These lines should be eliminated from the
list of measured equivalent widths.
The abundance of individual lines are showed in the file out2.txt (work.dir directory).

\section{Comparison with literature}

There are small differences in the parameters derived using this method by different authors.
\citet{santos00} derived surface gravities systematically higher than the
ones obtained by other authors \citep[e.g. ][]{gonzalez01} by 0.15 dex.
To solve this problem, \citet{santos04} adopted new log gf values for the
iron lines. Then the authors found a small average difference of +25 K in temperature
compared to the studies of \citet{gonzalez01} and \citet{laws03}
(57 stars in common). Also their log gs are +0.05 dex (on average) above
and the average differences in metallicity are between -0.10 and +0.10 dex.
\citet{ammler09} derived the fundamental parameters
of host stars with transiting planets, using the 2002 version of the code
MOOG and ATLAS model atmospheres.
However for the TrES and HAT objects, the abundance determinations of \citet{sozzetti07}
and \citet{sozzetti09} are systematically lower than those
derived in their work by {$\sim$0.10 dex}, using essentially the same method.
The authors also mention differences in temperature ($\sim$100 K) and
gravity ($\sim$0.15 dex) due to possible sistematic tendences between the works,
and they note that the origin of the discrepancies in abundances is commonly
unidentified.

We compare the parameters derived using {FUNDPAR} and those from
literature \citep{gonzalez97,gonzalez98,gonzalez99,santos00,gonzalez-laws00,
gonzalez01,laws-gonzalez01,laws03}. We selected these works in particular
because they present the Fe equivalent widths: we use the same values with FUNDPAR.
In the Table \ref{tabla.pars} we show the T$_{eff}$, $log g$, [FeI/H] and $\xi$ obtained.
Some stars in the table are listed twice (identified with a subscript) because we
have different equivalent width sources from literature. The sources of the equivalent
widths are listed in the Table \ref{tabla.pars} as R1,...,R8.

\begin{figure*}
\center
\includegraphics[width=120mm]{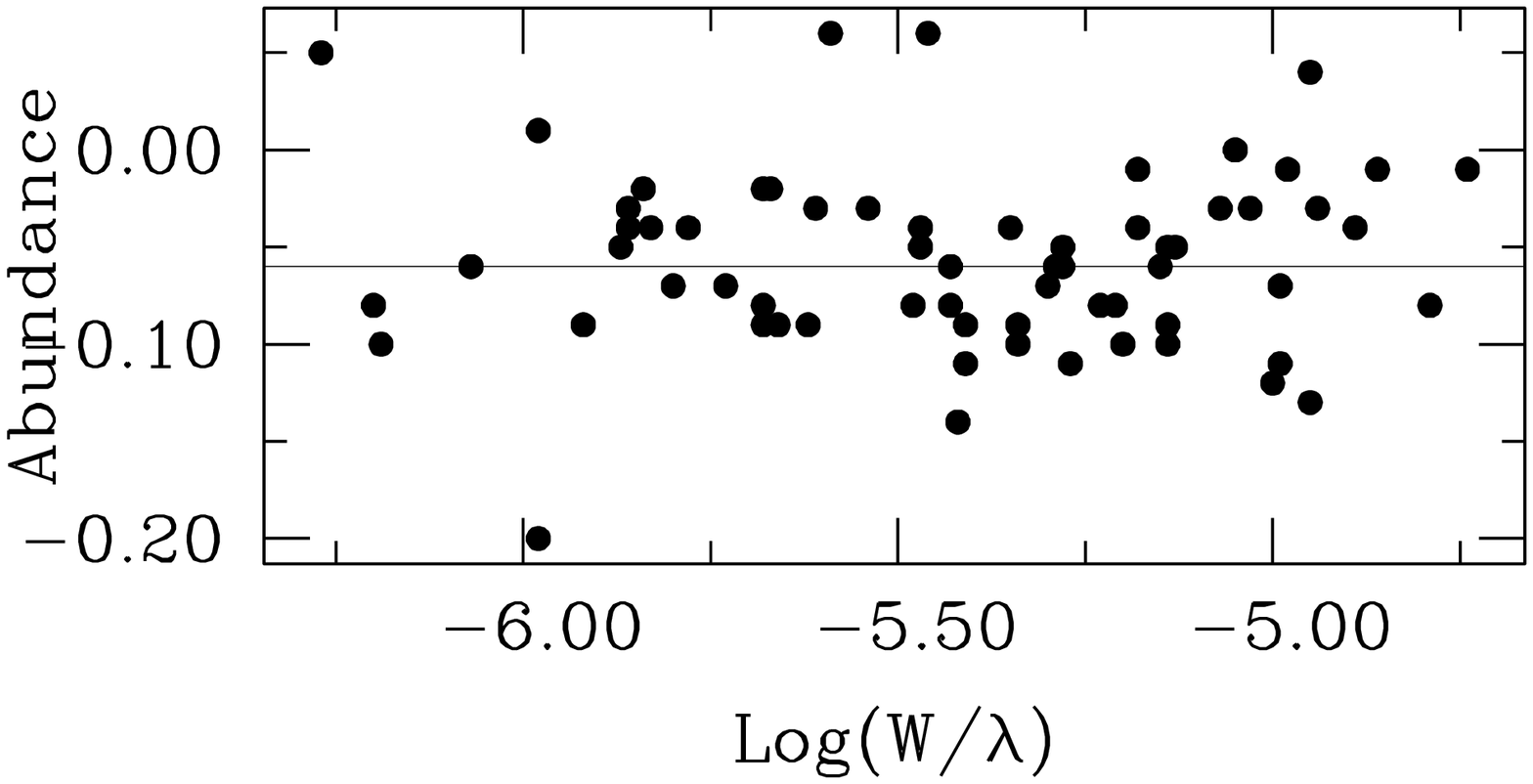}
\vskip -4.5in
\includegraphics[width=120mm]{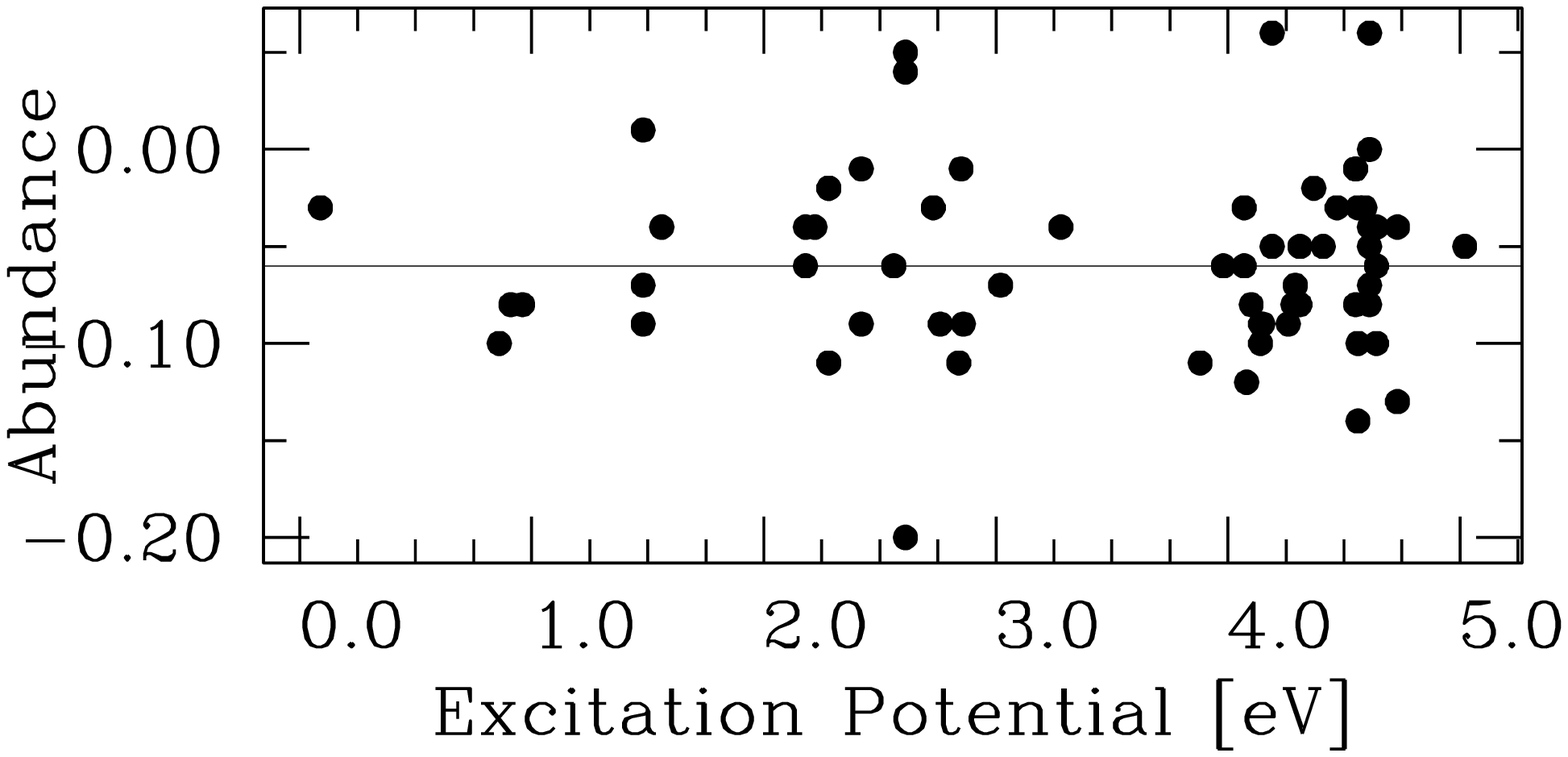}
\caption{Plots of abundance vs. excitation potential and abundance vs.
{$log_{10}(W/\lambda)$} (logarithm of the reduced equivalent width) for
HD 106252.}
\label{fig2}
\end{figure*}

\begin{figure*}
\center
\includegraphics[width=60mm]{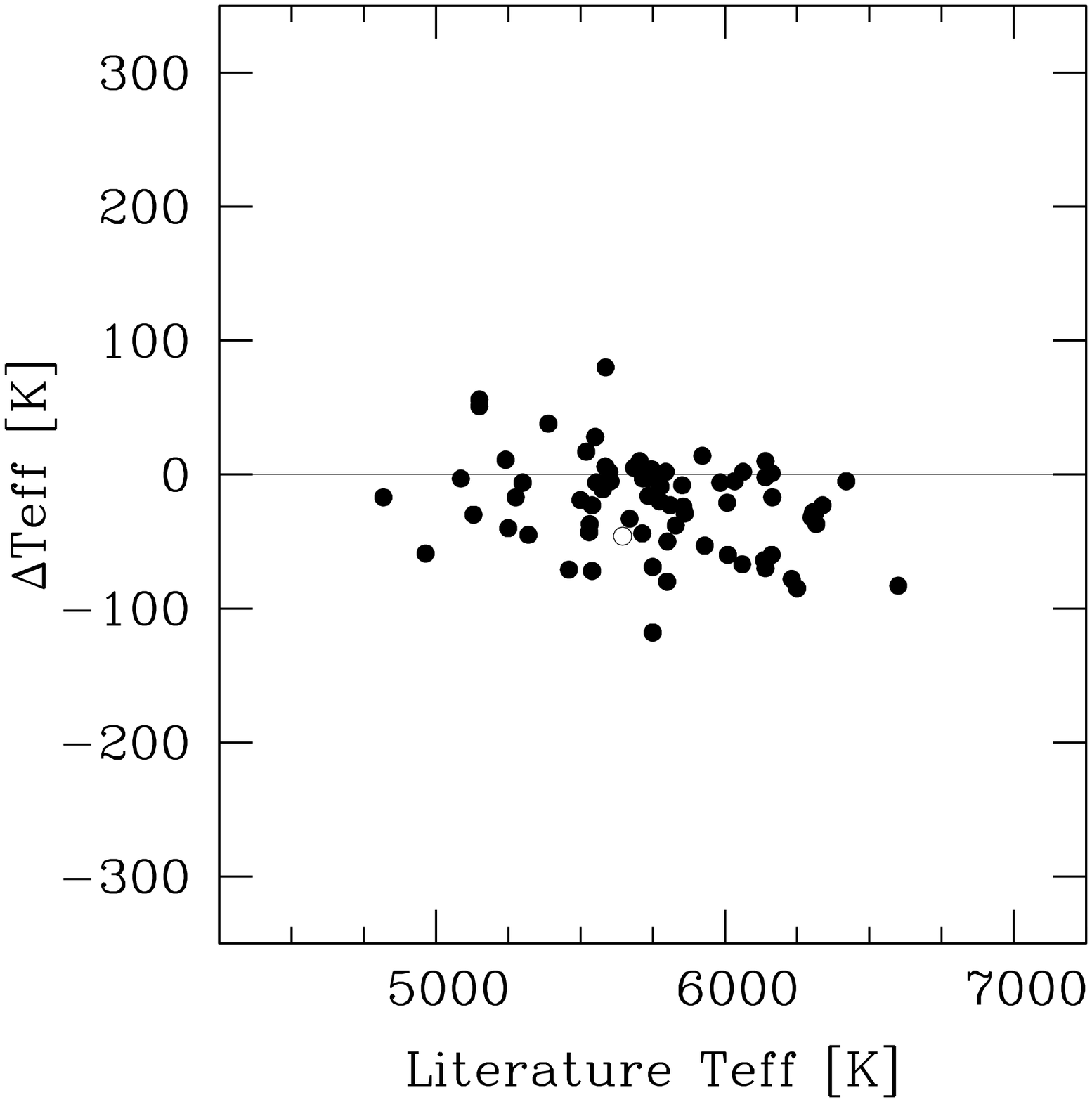}
\includegraphics[width=60mm]{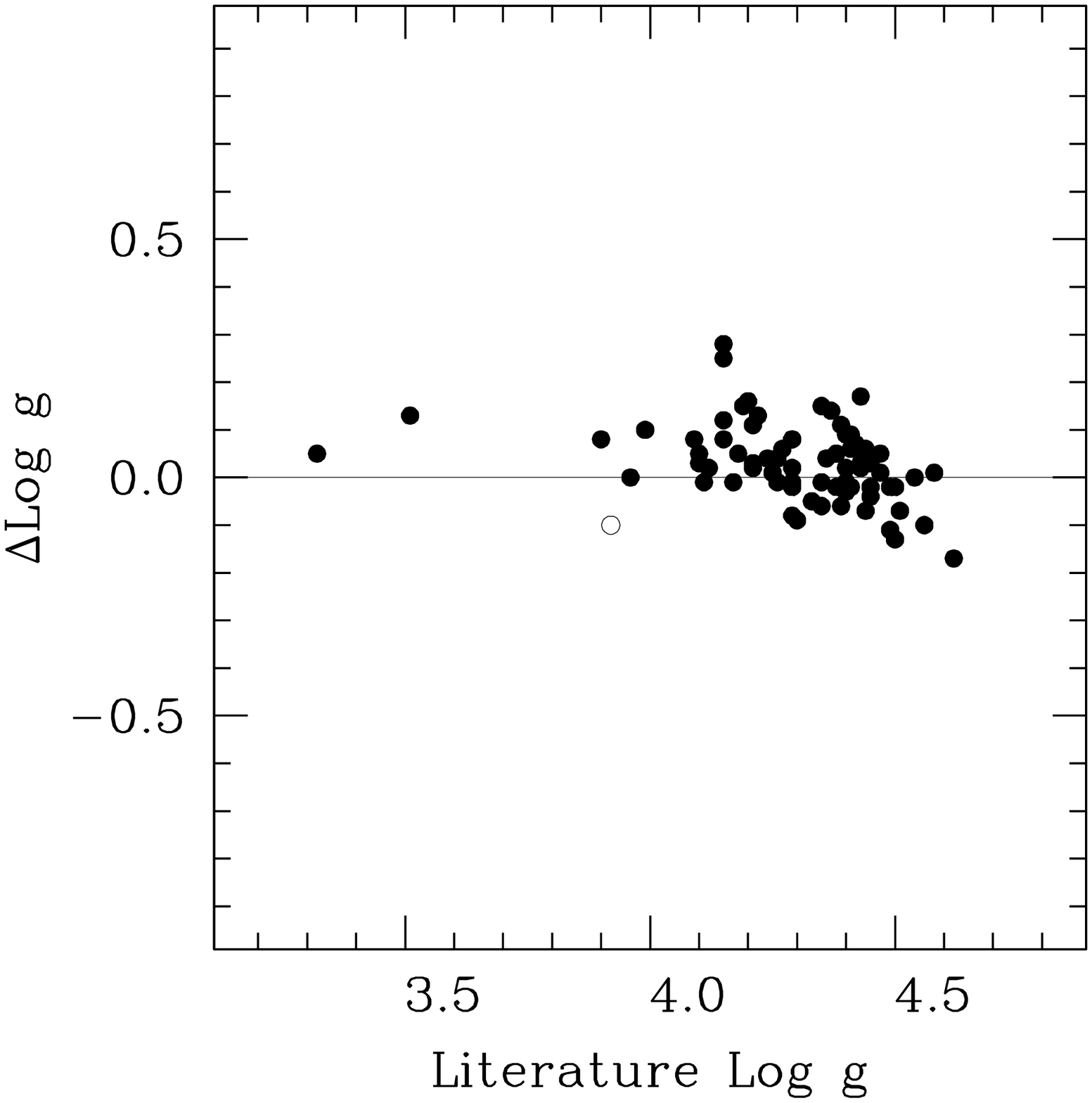}
\includegraphics[width=60mm]{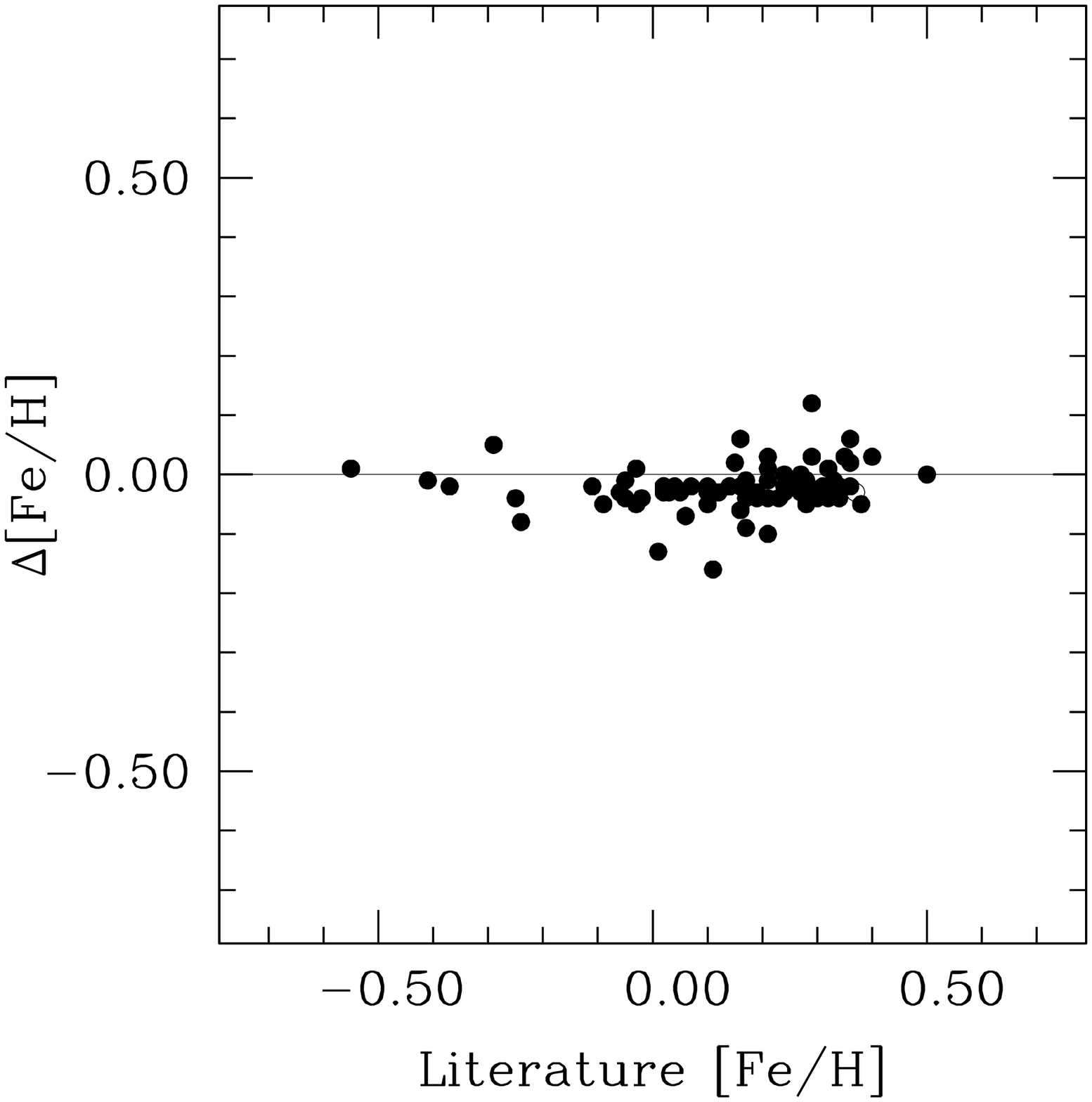}
\includegraphics[width=60mm]{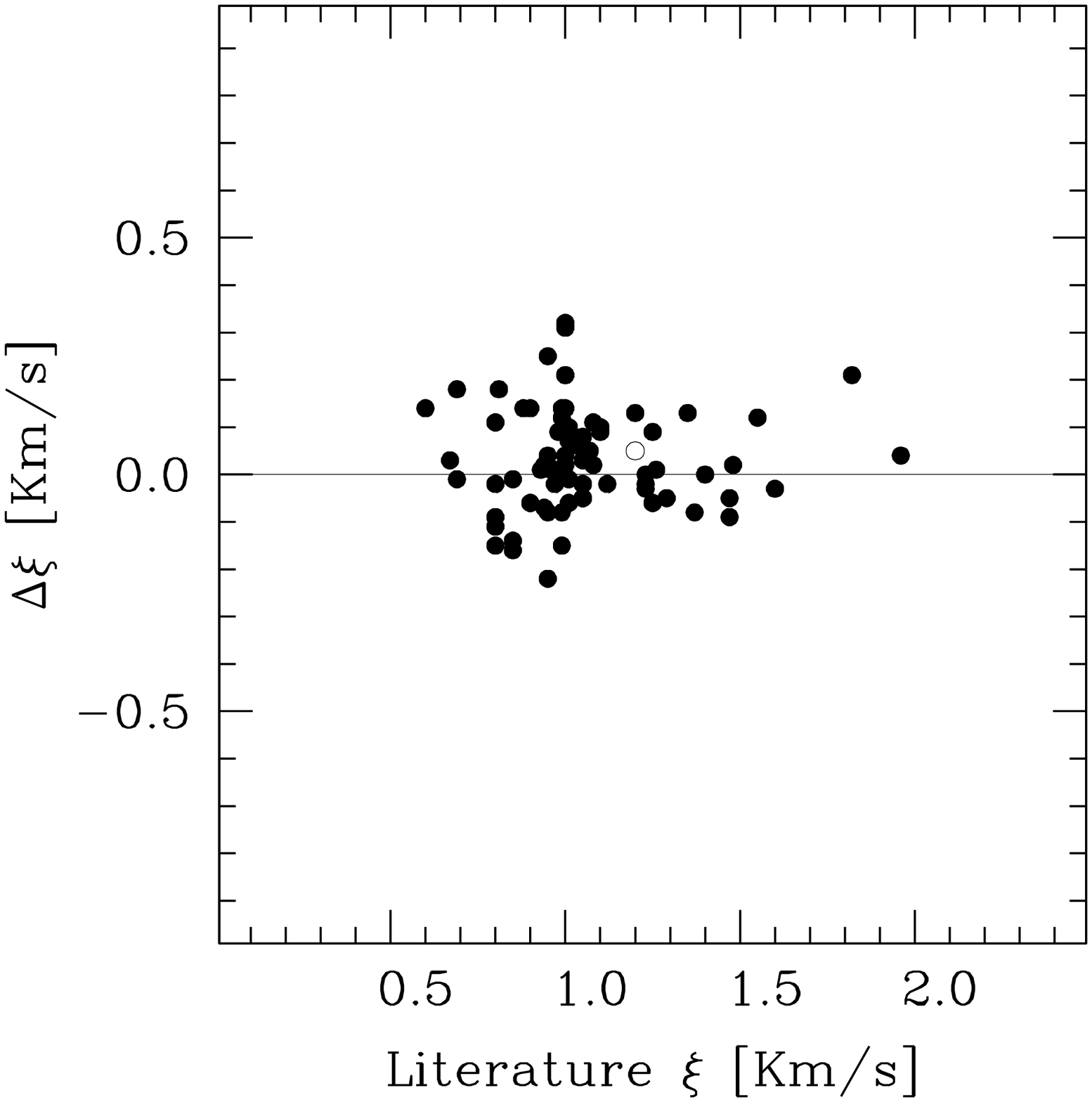}
\caption{
Difference between FUNDPAR parameters and literature (y axis) vs. literature
(x axis). The panels shows separately the effective temperature, gravity,
metallicity and $\xi$.
Filled and empty points correspond to solutions with {$\chi^2<$1} and
{$\chi^2>$1}, respectively.}
\label{fig1}
\end{figure*}

We show in the Figure \ref{fig2} an example of abundance vs. excitation potential and
abundance vs. {$log_{10}(W/\lambda)$} (logarithm of the reduced equivalent width) for
HD 106252.
In the Figure \ref{fig1} we present the difference between FUNDPAR parameters
and literature (y axis, $\Delta$=FUNDPAR-literature) vs. literature (x axis).
The empty point in the Figure \ref{fig1} correspond to the star HD 38529, for which their
slopes, metallicities and $\chi^2$ values indicate that this object should be taken with
caution. The derived parameters of this star are based only on 3 FeII lines (and 24 FeI lines).
In literature, some values have been rounded in T$_{eff}$, log g and $\xi$ to within
50 K or 10 K, 0.05 dex and 0.1 km/s, respectively
\citep[see, for example,][]{gonzalez97,gonzalez98,gonzalez99}.
There is a good agreement between {FUNDPAR} parameters and previous works from literature
within the errors, which is logic taking into account that {FUNDPAR} use a very similar method.
However the Figure shows that there is a dispersion in the values of the parameters 
(particularly $log g$ and $\xi$) and probably a sistematic tendence for the metallicity
($\sim$0.01 dex below literature values, see next discussion).
The values of $\Delta$T$_{eff}$ and $\Delta$log g also seems to slightly decrease
with T$_{eff}$ and log g, respectively.
The median of the differences for the fundamental parameters compared to literature are 24 K,
0.06 dex, 0.03 dex and 0.08 km/s, corresponding to T$_{eff}$, $log g$, [Fe/H] and $\xi$, respectively.
The higher differences in the parameters are 118 K (16 Cyg B), 0.30 dex (HD 27442), 0.16 dex
(16 Cyg B) and 0.31 km/s (47 Uma), corresponding to T$_{eff}$, $log g$, [Fe/H] and $\xi$, respectively.

Now we discuss the possible origin of the dispersions and probable tendences
observed in the Figure \ref{fig1}.
We tested FUNDPAR modifying significatively the values of the weights {w$_{1}$,...,w$_{4}$}
in the function $\chi^2$ and verified that the difference in [Fe/H] with literature and the
dispersion in the other parameters changes very slightly.
Then the weights {w$_{1}$,...,w$_{4}$} do not seem to be the cause.
We use a method similar to literature, however it is not totally identical.
FUNDPAR use different Kurucz model atmospheres \citep{castelli-kurucz03}
than those used in literature (most of them are previous to the creation of the ODFNEW models). 
The code use model atmospheres derived
by ATLAS using ODFNEW opacities and solar abundances from \citet{grevesse-sauval98} instead
of \citet{anders-grevesse89}. The new models present differences compared to older Kurucz
models \citep{kurucz90} such as the solar abundances, the replacement of TiO and H$_2$0 
molecular lines, some HI quasi-molecular absorptions, etc. taken into account in the NEWODF opacities. 
Preliminar improvements are the U-B and u-b color indices for {T$_{eff}<$6750 K}, all color
indices for cooler stars, and the better modeling for the upper layers of cool and giant stars
\citep{castelli-kurucz03}.
In this example, the model atmospheres are computed with convection (mixing-lenght parameter$=$1.25), 
overshooting (W=1) and damping in the \"Unsold approximation but multiplied by a factor as suggested
by \citet{blackwell95}.
The use of the fourth condition within the function $\chi^2$ is (possibly) another
difference with previous studies. Literature works surely take this into account, however it is
not totally clear for us how. Finally, we use the MOOG 2009 version\footnote{http://verdi.as.utexas.edu/moog.html}
of the program instead of the 2002 version used in literature, although we expect almost the
same abundance values from both versions.

These differences, at least in part, produce the slightly disimilar values showed in the Figure
\ref{fig1}, which suggest that FUNDPAR use probably a different metallicity scale than used in
literature. A complete comparison require the exact knowledge of all involved details used in the
literature calculation. Our intention is to clearly present all the asumptions used in FUNDPAR,
in the model atmospheres and within the code.

\begin{figure*}
\center
\includegraphics[width=60mm]{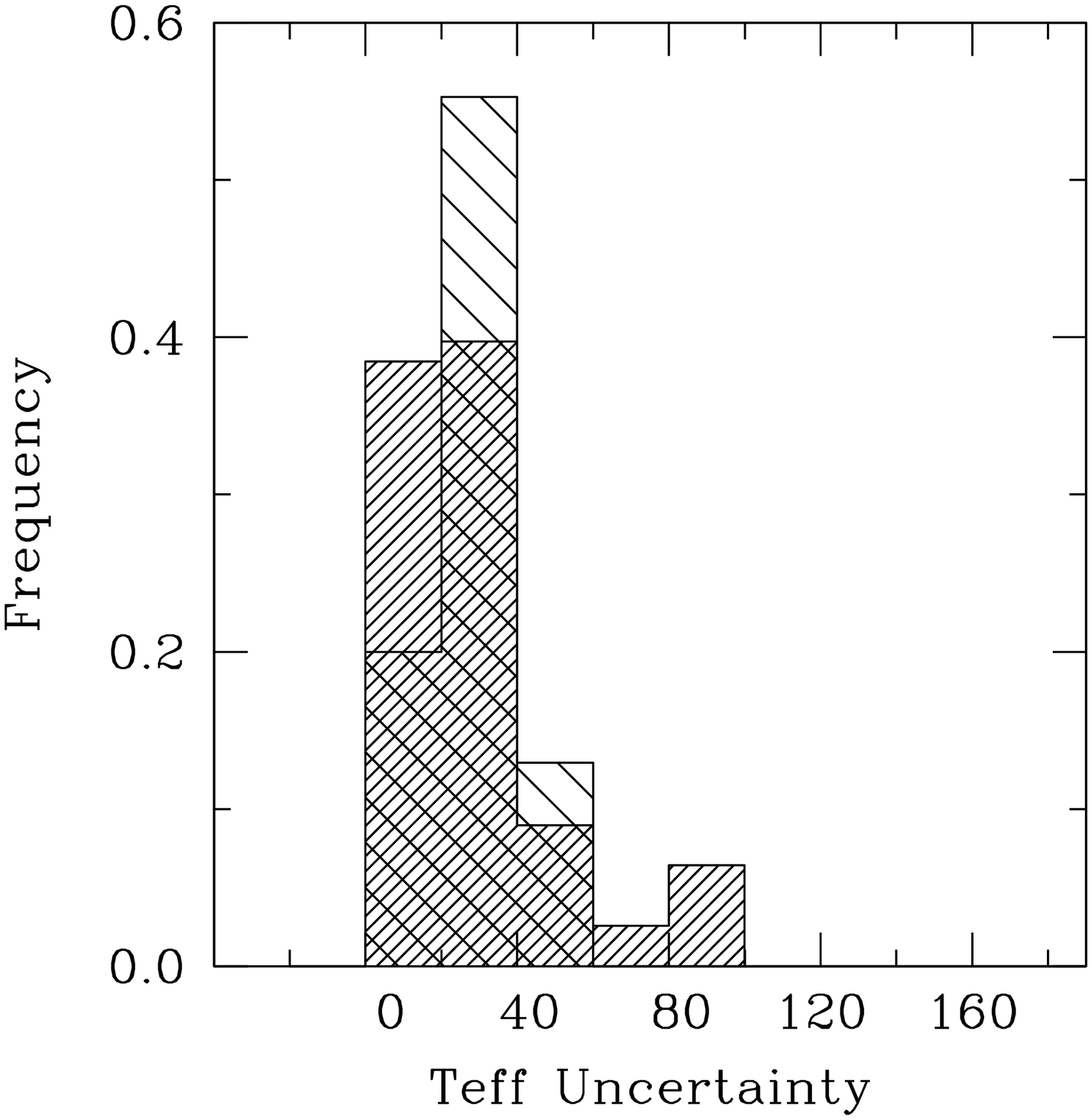}
\includegraphics[width=60mm]{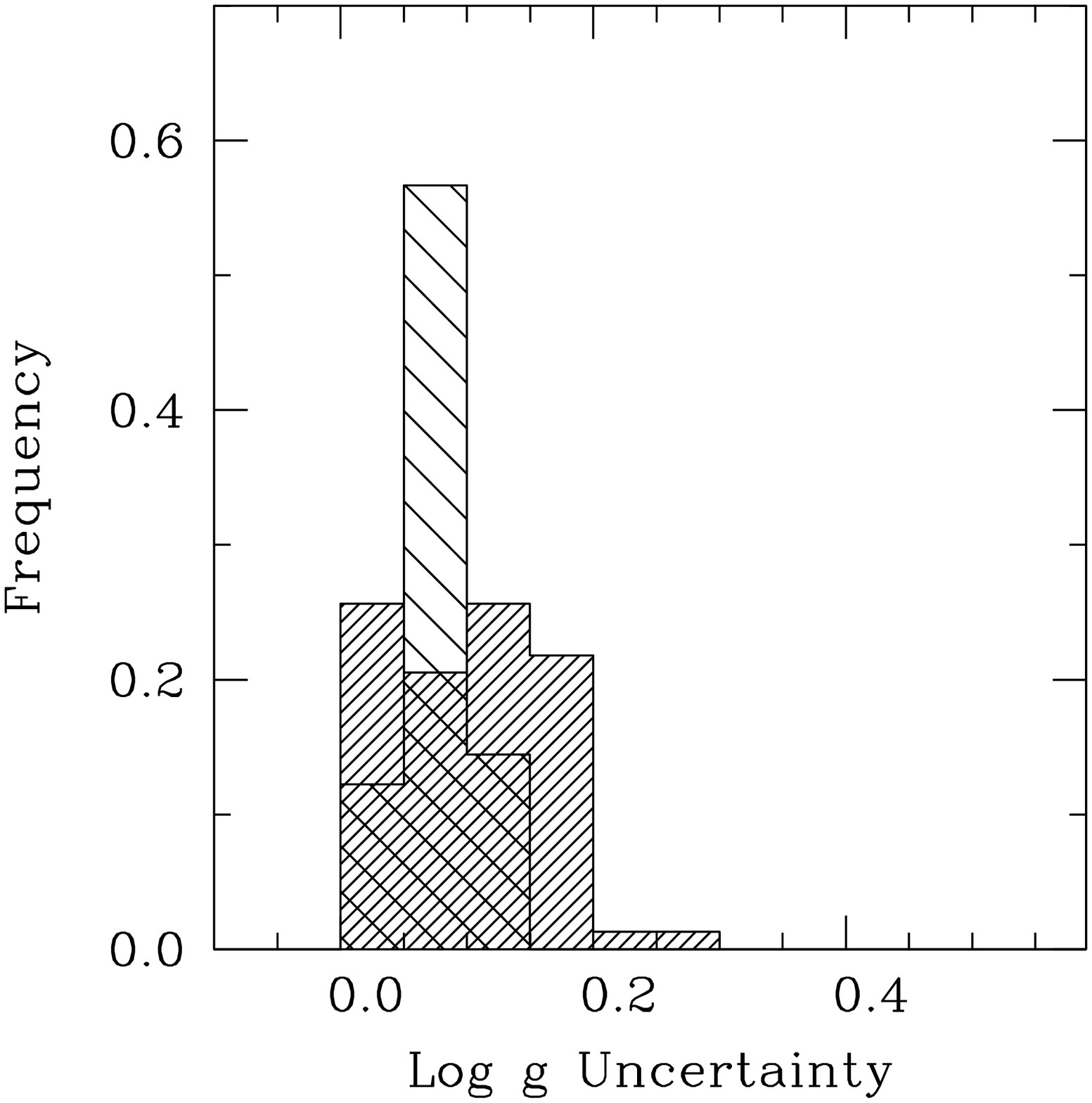}
\includegraphics[width=60mm]{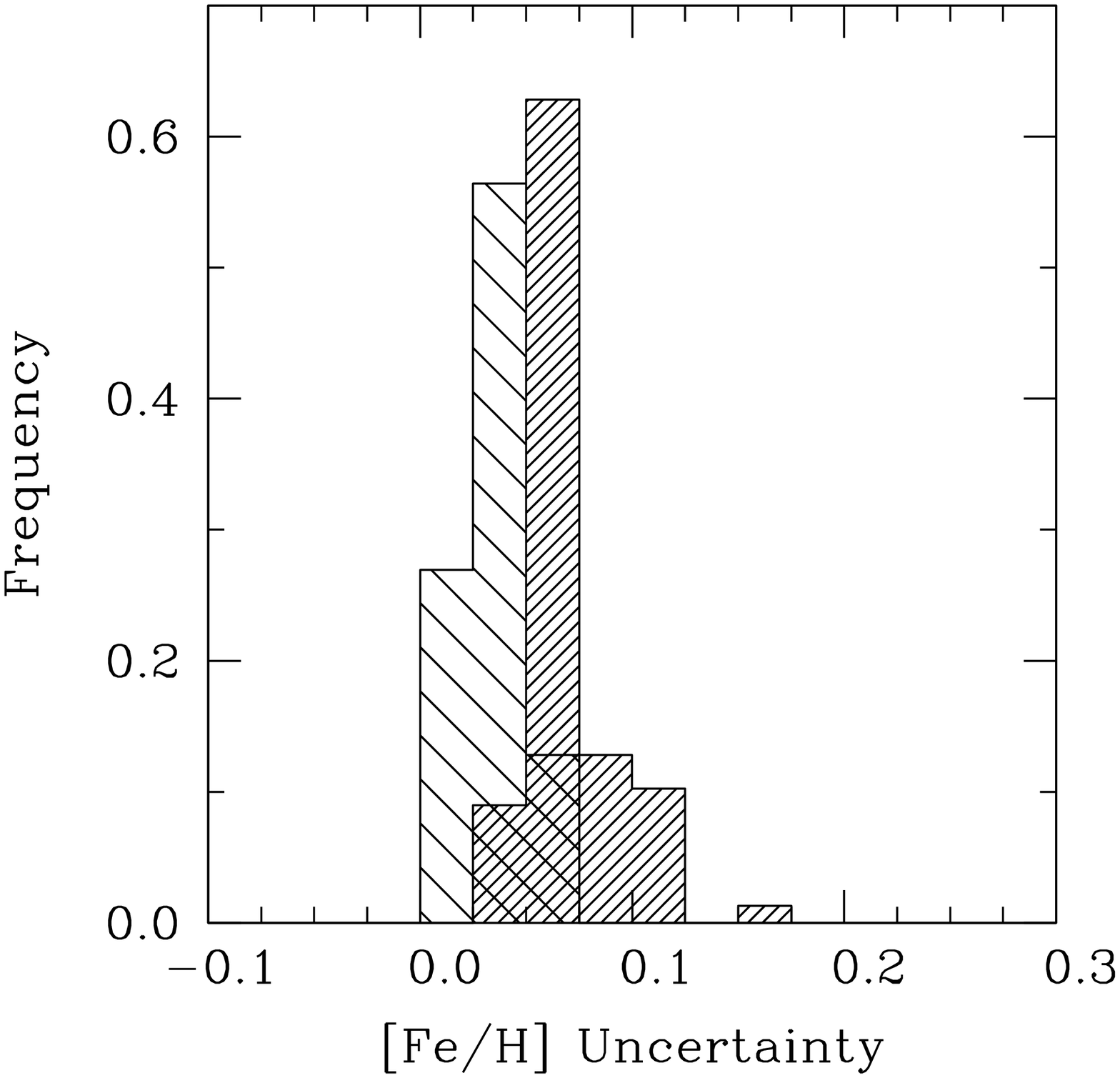}
\includegraphics[width=60mm]{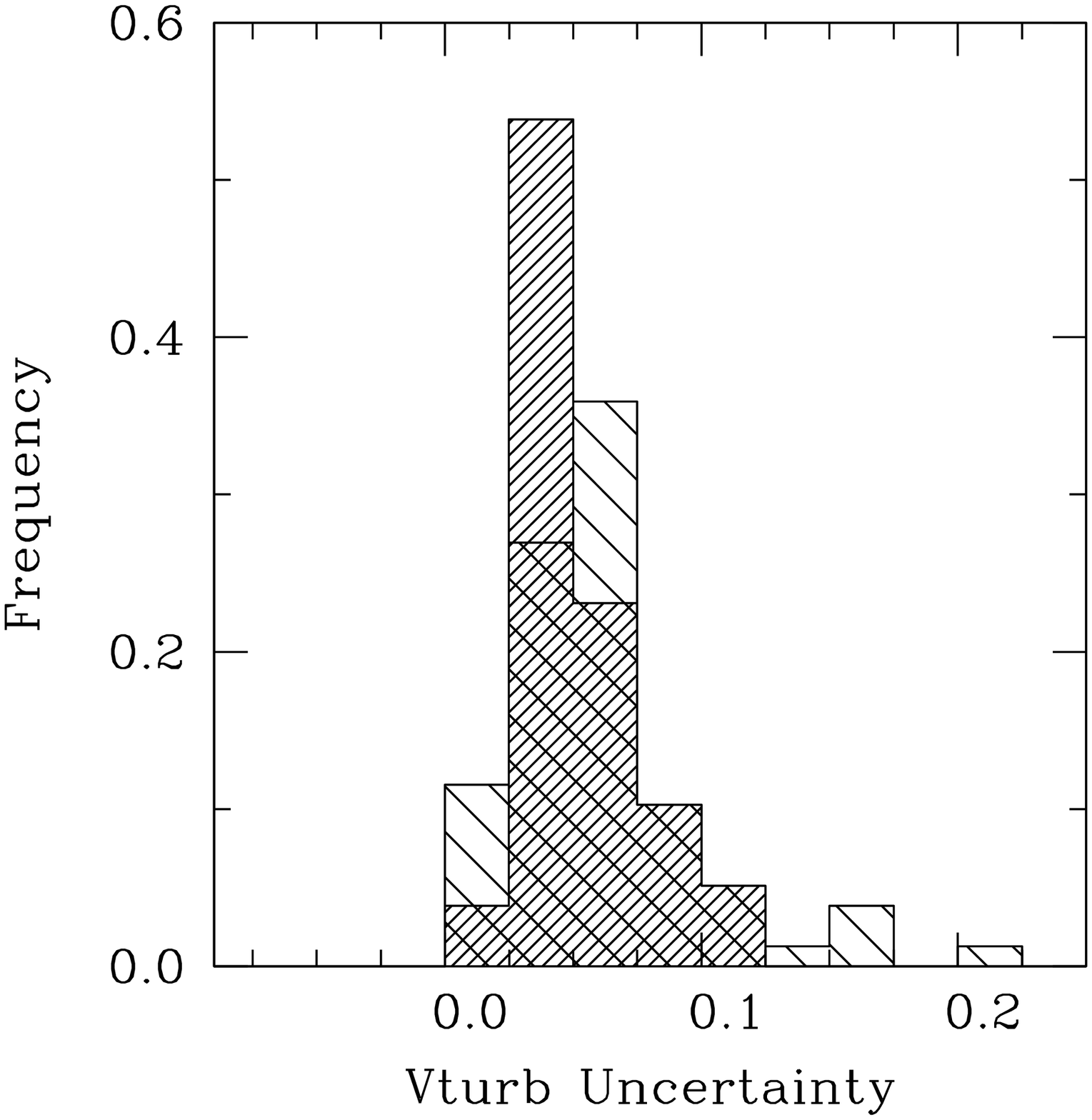}
\caption{Histogram distributions of the uncertainties in T$_{eff}$, $log g$, [Fe/H] and $\xi$
derived with FUNDPAR. The densely and slightly shaded histograms correspond to uncertaities
derived following \citet{gonzalez-vanture98} and using the $\chi^2$ function, respectively.}
\label{histog.uncer}
\end{figure*}

In The Figure \ref{histog.uncer} we show the histogram distributions of the uncertainties
derived in T$_{eff}$, $log g$, [Fe/H] and $\xi$.
The densely and slightly shaded histograms correspond to uncertaities derived following \citet{gonzalez-vanture98}
and using the $\chi^2$ function, respectively.
Some distributions present a peak in a common uncertainty value, such as $\sim$0.05 dex in the [Fe/H]
distribution and $\sim$30 K in the distribution of effective temperature.
We see that the errors derived using both methods are comparable.

\begin{figure}
\center
\includegraphics[width=60mm]{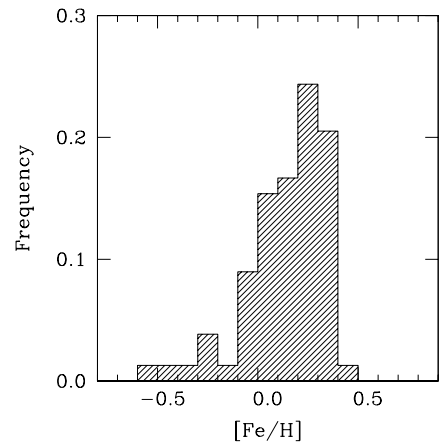}
\caption{Metallicity distribution of the sample stars derived by FUNDPAR.}
\label{histog.feh}
\end{figure}

The metallicities presented in the Table \ref{tabla.pars} correspond to a group
of exoplanet host stars. The median of the group is 0.17 dex with a dispersion of 0.22 dex. 
In the Figure \ref{histog.feh} we present the histogram of the metallicity distribution.
Then, as an example of practical use of {FUNDPAR}, we verified the
metal-rich nature of main sequence stars with low mass companions, a fact known from the
literature \citep[see, for example,][]{gonzalez97,santos00}.

\section{Concluding remarks}

We have implemented a fortran algorithm available from the web
that estimate fundamental parameters of solar type stars, requiring only
the measure of Fe equivalent widths.
The final solution should verify the three conditions of the standard method:
{[FeI/H]$=$[FeII/H]} (i.e. ionization equilibrium), independence of the metallicity
with the excitation potential (i.e. excitation equilibrium) and with respect to
the equivalent widths. 
We also add another condition: the input metallicity used in the model atmosphere
should be similar to the resulting metallicity from the equivalent widths.
We taken into account these conditions in one variable called $\chi^2$,
adopting an expression which include the weights {w$_{1}$,...,w$_{4}$}.
FUNDPAR use Kurucz model atmospheres with the NEWODF opacities \citep{castelli-kurucz03},
solar-scaled abundances from \citet{grevesse-sauval98} and the 2009 version of the MOOG program.
Different details could be selected, such as the mixing-lenght parameter, the
overshooting and the damping of the lines, for instance.
We have planed a new version that include the option of use the WIDTH9 program
instead of MOOG deriving abundances from equivalent widths.

The code include the derivation of the uncertainty in the 4 parameters
following the criteria of Gonzalez \& Vanture (1998) and another uncertainty
estimation using the $\chi^2$ function. We verified the metal-rich nature of a group of
exoplanet host stars. The parameters derived with {FUNDPAR} are in agreement with
previous works in literature.

\begin{longtable}{lrrrrrrrrr}
\caption{Fundamental parameters and dispersions derived with FUNDPAR.
References for the equivalent widths (last column): R1: \citet{gonzalez99},
R2: \citet{gonzalez98},
R3: \citet{laws-gonzalez01},
R4: \citet{gonzalez01},
R5: \citet{laws03},
R6: \citet{santos00},
R7: \citet{gonzalez00},
R8: \citet{gonzalez97}
}
\\
\hline \hline
Star  & T$_{eff}$ & log g & [Fe/H] & $\xi$ & eTeff & elog g & e[Fe/H] & e$\xi$ & Ref\\
      & [K]       & [dex] & [dex] & [km/s] & [K]   & [dex]  &  [dex]  & [km/s] \\
\hline
\endfirsthead
\caption{Continued.}\\
\hline \hline
Star  & T$_{eff}$ & log g & [Fe/H] & $\xi$ & eTeff & elog g & e[Fe/H] & e$\xi$ & Ref\\
      & [K]       & [dex] & [dex] & [km/s] & [K]   & [dex]  &  [dex]  & [km/s] \\
\hline
\endhead
\hline
\endfoot
14 Her           &    5294&    4.33&    0.50&    0.71&     40.60&    0.19&    0.07&    0.05& R1      \\
16 Cyg A$_1$     &    5704&    4.29&   -0.01&    1.32&     90.75&    0.19&    0.11&    0.10& R2      \\
16 Cyg A$_2$     &    5749&    4.23&    0.07&    1.02&     15.58&    0.04&    0.06&    0.03& R3      \\
16 Cyg B$_1$     &    5632&    4.36&   -0.05&    1.21&     83.68&    0.25&    0.10&    0.09& R2      \\
16 Cyg B$_2$     &    5690&    4.30&    0.05&    0.91&     14.16&    0.04&    0.06&    0.03& R3      \\
47 Uma           &    5720&    4.26&   -0.12&    1.31&     86.63&    0.15&    0.10&    0.08& R2      \\
51 Peg$_1$       &    5681&    4.37&    0.11&    1.21&     62.21&    0.11&    0.08&    0.08& R2      \\
51 Peg$_2$       &    5797&    4.39&    0.22&    1.13&     16.33&    0.04&    0.04&    0.03& R4      \\
70 Vir           &    5481&    3.98&   -0.08&    1.10&     92.16&    0.15&    0.09&    0.07& R2      \\
BD-10 3166       &    5275&    4.36&    0.32&    0.69&     42.70&    0.05&    0.09&    0.05& R4      \\
HD 106252        &    5844&    4.50&   -0.06&    0.98&     12.23&    0.12&    0.05&    0.03& R5      \\
HD 10697         &    5600&    3.96&    0.10&    1.20&     28.41&    0.03&    0.05&    0.04& R4      \\
HD 108147        &    6279&    4.59&    0.19&    1.11&     41.51&    0.12&    0.11&    0.09& R5      \\
HD 114783        &    5100&    4.48&    0.08&    0.87&     15.70&    0.10&    0.07&    0.04& R5      \\
HD 117176        &    5487&    4.09&   -0.06&    1.00&     10.86&    0.10&    0.06&    0.02& R5      \\
HD 1237$_1$      &    5537&    4.50&    0.22&    1.19&     30.76&    0.15&    0.07&    0.05& R4      \\
HD 1237$_2$      &    5468&    4.50&    0.05&    1.38&     23.27&    0.18&    0.07&    0.04& R6     \\
HD 12661$_1$     &    5670&    4.41&    0.38&    0.91&     19.18&    0.06&    0.05&    0.03& R4      \\
HD 12661$_2$     &    5719&    4.48&    0.38&    0.84&     24.28&    0.04&    0.05&    0.04& R4      \\
HD 13445         &    5080&    4.43&   -0.30&    0.65&     19.69&    0.23&    0.07&    0.05& R6     \\
HD 134987        &    5719&    4.28&    0.33&    1.09&     22.44&    0.04&    0.07&    0.04& R4      \\
HD 136118        &    6153&    4.37&   -0.09&    2.00&     21.82&    0.15&    0.08&    0.11& R5      \\
HD 141937        &    5832&    4.50&    0.12&    0.97&     12.63&    0.10&    0.05&    0.03& R5      \\
HD 160691        &    5788&    4.49&    0.24&    1.12&     22.03&    0.08&    0.06&    0.04& R5      \\
HD 16141$_1$     &    5768&    4.24&    0.17&    1.10&     22.07&    0.04&    0.04&    0.04& R4      \\
HD 16141$_2$     &    5751&    4.28&    0.16&    1.00&     12.18&    0.05&    0.05&    0.03& R5      \\
HD 168443        &    5549&    4.13&    0.08&    1.04&     17.06&    0.02&    0.06&    0.03& R4      \\
HD 168746        &    5566&    4.43&   -0.09&    0.94&     10.51&    0.04&    0.05&    0.03& R5      \\
HD 169830$_1$    &    6268&    4.10&    0.20&    1.29&     22.14&    0.18&    0.05&    0.03& R6     \\
HD 169830$_2$    &    6283&    4.23&    0.16&    1.27&     24.27&    0.08&    0.06&    0.06& R5      \\
HD 177830        &    4801&    3.37&    0.38&    0.95&     55.16&    0.07&    0.10&    0.04& R4      \\
HD 187123        &    5792&    4.39&    0.14&    1.04&     26.26&    0.08&    0.04&    0.04& R1      \\
HD 190228        &    5259&    3.64&   -0.32&    1.13&     11.79&    0.11&    0.05&    0.02& R5      \\
HD 192263        &    4905&    4.38&   -0.02&    0.73&     19.76&    0.18&    0.08&    0.05& R4      \\
HD 195019a       &    5731&    4.17&    0.00&    1.20&     13.09&    0.13&    0.05&    0.03& R5      \\
HD 19994.2       &    6147&    4.35&    0.12&    2.03&     23.94&    0.10&    0.08&    0.09& R5      \\
HD 202206$_1$    &    5635&    4.55&    0.28&    0.88&     24.27&    0.14&    0.06&    0.04& R6     \\
HD 202206$_2$    &    5713&    4.45&    0.30&    1.07&     14.42&    0.04&    0.06&    0.03& R5      \\
HD 209458        &    6065&    4.43&    0.02&    1.14&     18.56&    0.13&    0.06&    0.04& R4      \\
HD 210277        &    5517&    4.34&    0.24&    0.84&     26.56&    0.12&    0.05&    0.04& R1      \\
HD 217107        &    5602&    4.42&    0.42&    0.87&     17.64&    0.03&    0.05&    0.03& R4      \\
HD 22049         &    5083&    4.47&   -0.14&    0.84&     14.29&    0.15&    0.06&    0.04& R5      \\
HD 222582        &    5719&    4.25&    0.00&    0.99&     18.90&    0.03&    0.04&    0.04& R4      \\
HD 27442         &    4859&    3.60&    0.38&    1.21&     45.85&    0.22&    0.15&    0.06& R5      \\
HD 28185         &    5637&    4.54&    0.21&    0.96&     13.79&    0.04&    0.06&    0.03& R5      \\
HD 33636         &    5877&    4.27&   -0.13&    0.95&     14.81&    0.10&    0.06&    0.04& R5      \\
HD 37124$_1$     &    5579&    4.60&   -0.39&    0.74&     19.83&    0.11&    0.09&    0.08& R5      \\
HD 37124$_2$     &    5495&    4.46&   -0.42&    0.71&     22.54&    0.04&    0.05&    0.07& R4      \\
HD 38529         &    5600&    3.82&    0.34&    1.25&     38.21&    0.05&    0.06&    0.03& R4      \\
HD 4203          &    5667&    4.43&    0.43&    1.08&     29.95&    0.06&    0.15&    0.06& R5      \\
HD 4208          &    5592&    4.33&   -0.29&    0.87&     14.74&    0.15&    0.06&    0.04& R5      \\
HD 46375$_1$     &    5210&    4.37&    0.24&    0.65&     35.26&    0.10&    0.06&    0.04& R4      \\
HD 46375$_2$     &    5252&    4.50&    0.26&    0.68&     21.15&    0.09&    0.10&    0.06& R5      \\
HD 50554         &    5978&    4.51&   -0.01&    1.10&     11.93&    0.10&    0.06&    0.03& R5      \\
HD 52265$_1$     &    6102&    4.28&    0.24&    1.20&     17.71&    0.06&    0.05&    0.03& R4      \\
HD 52265$_2$     &    6163&    4.37&    0.27&    1.21&     19.51&    0.10&    0.03&    0.03& R4      \\
HD 52265$_3$     &    5993&    4.21&    0.17&    1.24&     25.17&    0.18&    0.06&    0.04& R6     \\
HD 6434          &    5715&    4.32&   -0.54&    0.70&     23.11&    0.16&    0.11&    0.10& R5      \\
HD 68988         &    5936&    4.49&    0.32&    1.19&     19.40&    0.02&    0.09&    0.05& R5      \\
HD 75289$_1$     &    6138&    4.52&    0.27&    1.50&     20.96&    0.10&    0.07&    0.05& R7     \\
HD 75289$_2$     &    6070&    4.44&    0.23&    1.42&     30.93&    0.16&    0.06&    0.05& R6     \\
HD 75332         &    6277&    4.47&    0.22&    1.03&     24.73&    0.11&    0.05&    0.04& R4      \\
HD 82943$_1$     &    5950&    4.45&    0.28&    1.10&     25.37&    0.07&    0.05&    0.03& R6     \\
HD 82943$_2$     &    5987&    4.46&    0.24&    1.11&     17.85&    0.04&    0.05&    0.03& R5      \\
HD 83443$_1$     &    5389&    4.31&    0.33&    1.00&     27.40&    0.18&    0.08&    0.04& R6     \\
HD 83443$_2$     &    5427&    4.40&    0.34&    0.99&     22.07&    0.13&    0.07&    0.05& R5      \\
HD 8574          &    6029&    4.34&    0.00&    1.19&     13.29&    0.16&    0.05&    0.03& R5      \\
HD 89744         &    6315&    4.16&    0.27&    1.67&     28.06&    0.07&    0.04&    0.05& R4      \\
HD 92788         &    5755&    4.43&    0.29&    1.02&     26.37&    0.04&    0.05&    0.04& R4      \\
HD 95128         &    5832&    4.31&    0.02&    1.08&     13.83&    0.10&    0.05&    0.02& R5      \\
HR 810           &    6072&    4.48&    0.15&    1.23&     27.89&    0.09&    0.07&    0.08& R4      \\
$\nu$ And$_1$    &    6165&    4.21&    0.13&    1.40&     50.44&    0.03&    0.06&    0.07& R8     \\
$\nu$ And$_2$    &    6150&    4.14&    0.09&    1.48&     50.48&    0.03&    0.07&    0.07& R7     \\
$\rho$ Cnc$_1$   &    5201&    4.27&    0.32&    0.78&     99.87&    0.20&    0.11&    0.10& R2      \\
$\rho$ Cnc$_2$   &    5206&    4.40&    0.41&    0.69&     27.42&    0.18&    0.07&    0.04& R7     \\
$\rho$ Crb       &    5751&    4.15&   -0.24&    1.33&     28.85&    0.05&    0.04&    0.05& R2      \\
$\tau$ Boo$_1$   &    6517&    4.37&    0.30&    1.57&     60.99&    0.04&    0.06&    0.08& R8     \\
$\tau$ Boo$_2$   &    6415&    4.23&    0.30&    1.34&     49.61&    0.06&    0.08&    0.07& R7     \\
\label{tabla.pars}
\end{longtable}


\twocolumn
\begin{acknowledgements}
The authors thank R.L. Kurucz, C. Sneden, L. Sbordone, P. Bonifacio
and F. Castelli for making their codes available to them.
We also thank the anonymous referee whose suggestions improved the paper.
This work was partially supported by a grant PIP 1153 from 
Consejo Nacional de Investigaciones Cient\'{i}ficas y T\'{e}cnicas de
la Rep\'{u}blica Argentina.
\end{acknowledgements}

\end{document}